\documentclass[aps,pra,showpacs,twocolumn]{revtex4}
\usepackage{mathrsfs}
\usepackage{amssymb}
\usepackage{amsmath}
\usepackage{bm}
\usepackage{graphics}
\usepackage{natbib}
\usepackage{color}

\begin{document}
\title{Topological properties of the dimerized Kitaev chain with long-range hopping and pairing terms}

\author{Xue-Si Li$^1$}
\author{Jia-Rui Li$^1$}
\author{Shu-Feng Zhang$^2$}
\author{Lian-Lian Zhang$^1$}
\author{Wei-Jiang Gong$^1$}\email{gwj@mail.neu.edu.cn}

\affiliation{1. College of Sciences, Northeastern University, Shenyang 110819, China\\
2. School of Physics and Technology, University of Jinan, Jinan 250022, China}
\date{\today}

\begin{abstract}
We investigate the topological properties of a dimerized Kitaev chain with long-range interactions, including the intercell hopping and superconducting pairing terms. It is found that even only when the intercell hopping term appears, the size of the energy gap, the proportion of topological phases, and the topological phase transition can be modulated. The notable result is that they lead to a new Kitaev-like phase featured by the twofold-degenerated Majorana zero-energy edge states. Next in the presence of the intercell superconducting pairing term, this kind of Majorana phase can be magnified. This work provides new proposals to realize the twofold degenerated Majorana modes based on the intercell hopping and superconducting pairing terms of the dimerized Kitaev chain.

\end{abstract}
\pacs{11.30.Er, 03.65.Vf, 73.21.Cd}
\maketitle

\bigskip

\section{Introduction}
Majorana fermions, the particles whose antiparticles are identical to themselves, have received a great deal of attention in the field of high energy physics\cite{high energy1,high energy2,high energy3}. As their counterpart in the field of condensed matter physics, the Majorana zero-energy modes (MZMs) have been reported to exist at the edges of the topological superconductors\cite{TS1,TS2,TS3,TS4}. Since the MZMs obey non-Abelian statistics\cite{non-Abelian}, they have the robustness against environmental disturbance, and can manipulate the quantum information through the topological braiding. These valuable properties promise the MZMs to be the appropriate candidates for the fault-tolerant quantum computation\cite{fault-tolerant}. Therefore, they have been investigated extensively in both theories and experiments over the past years\cite{MZMs1,MZMs2,MZMs3,MZMs4}.
\par
As reported in the previous works, the MZMs can be hosted by lots of real topological-superconductor (TS) systems and are allowed to appear at the ends of the one-dimensional $p$-wave superconductor (i.e., the Kitaev chain)\cite{1D-SC1,1D-SC2} and the vortex core of the two-dimensional $p_{x}+ip_{y}$ superconductor\cite{vortex1,vortex2,vortex3,vortex4}. As for the $p$-wave superconductor, it is usually realized by placing the semiconductor nanowire with strong spin-orbit coupling on the surface of the $s$-wave superconductor, under the condition of the perpendicular Zeeman field\cite{Oreg,Pientka,Delft}. And the later can be realized when the two-dimensional topological insulator adheres to the $s$-wave superconductors \cite{2D-TS1,2D-TS2}. According to the characteristics of the MZMs, the above systems are called the the D-class superconductors\cite{D1,D2}, due to the breaking of the time-reversal symmetry. In resent years, researchers have begun to pay attention to the MZMs with other symmetries. One example is the $\mathrm{DIII}$-class TS, in which the MZM appears in the form of Kramers doublet because of time-reversal symmetry\cite{Zhang2,Fuliang2,Deng,Nakosai,Wong,Zhang,Nagaosa2,Gong}. It has been found that the MZM doublet causes the period of supercurrent to be dependent on the fermion parity of the Josephson junction\cite{Liuxj,Qixl}. The other is the BDI-class TS\cite{BDI,BDI1}, which can be built with superconductors coupled to the AIII topological insulators which show the quantum anomalous Hall effect\cite{BDI2,BDI3,BDI4}. It shows that the zero-bias conductance of the normal metal(N)-TS junction shows zero or $4e^2/h$ value depending on the phase of the tunnel
coupling between the normal metal and the TS\cite{BDI4}.
\par
Considering the above research progresses, one should notice that the one-dimensional TS systems are the significant candidates for the realization of MZMs, and their types directly induce the different properties of the MZMs. Meanwhile, the theoretical schemes play important roles for searching the MZMs. Following these facts, various Kitaev-like models have been proposed, including the dimerized Kitaev chain that consists of the Su-Schrieffer-Heeger (SSH) model and the Kitaev model\cite{dimerized kitaev1, dimerized kitaev2,dimerized kitaev3}, and the extended Kitaev chain with longer-range hopping and superconducting pairing (SP) terms\cite{long range1,long range2,long range3}. In addition to realization of the MZMs, these models also exhibit abundant topological phases, distinguished by the topological numbers and the zero-bias differential conductance. One can then ascertain that it is meaningful to explore new Kitaev-like systems, to clarify the appearance of the MZMs.
\par
Inspired by the existed researches, we would like to consider a dimerized Kitaev chain with long-range interactions for the intercell hopping or SP terms to investigate the topological phases and zero-energy modes. According to the symmetry classification, the system belongs to the BDI class, thus the topological phase can be characterized by the winding numbers in the $k$ space. By plotting the phase diagram of the winding numbers and the energy spectrum, we find that the intercell hopping terms induce a new Kitaev-like phase that exhibits the twofold-degenerated MZMs at each end of the chain. Next the long-range intercell SPs are incorporated, the interplay between these two kinds of long-range interactions can magnify the region of the new Kitaev-like phase. Furthermore, in the presence of weak disorder, the two-fold degenerated MZMs can be kept.

The rest of the paper is organized as follows. In Sec. II, we give the theoretical model, derive its energy bands, discuss its symmetry, and calculate the winding numbers. In Sec. III, the topological phase diagrams, the energy spectra and the Andreev-reflection conductance are investigated. In Sec. IV, we present a brief summary.
\begin{figure}[htp]
\centering \scalebox{0.23}{\includegraphics{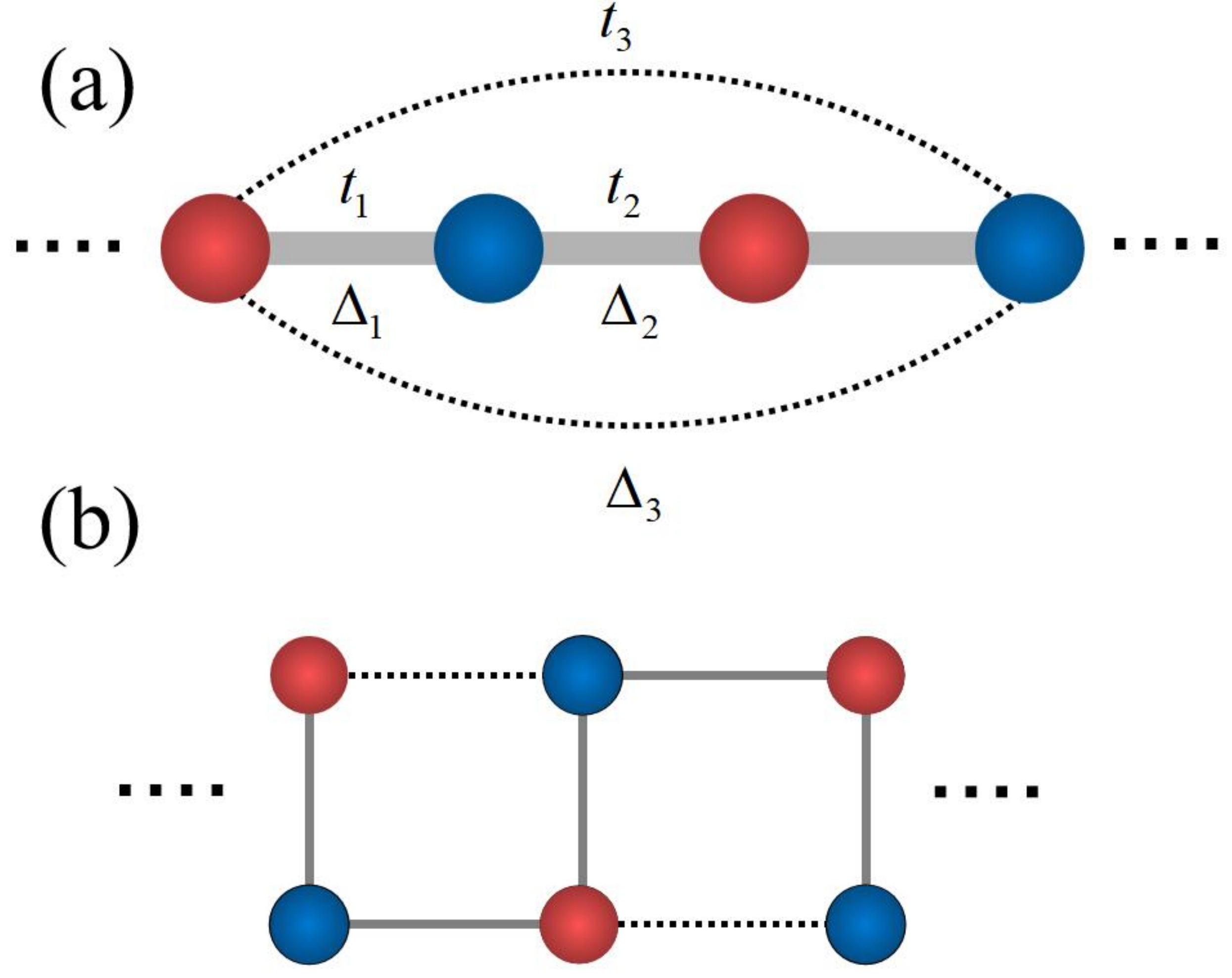}}
\caption{ (a) Schematic of the dimerized Kitaev chain with long-range intercell hopping and superconducting pairing terms (i.e., $t_3$ and $\Delta_3$). (b) The equivalent ladder lattice of the hybrid model in (a).\label{fig1}}
\end{figure}
\section{Theoretical model}
The dimerized Kitaev chain that we consider is illustrated in Fig.1(a), in which the intercell hopping and SP terms are also presented. Its Hamiltonian is written as
\begin{eqnarray}
H&=&H_{0}+H_{I}.
\end{eqnarray}
$H_{0}$ represents the Hamiltonian of the leading part of the dimerized Kitaev chain, i.e.,
\begin{eqnarray}
H_{0}&=&\sum_{j=1}^{\mathcal{N}-1}(t_{1}c_{B,j}^{\dag}c_{A,j}+t_{2}c_{A,j+1}^{\dag}c_{B,j}+\mathrm{h.c.})\notag\\
&+&\sum_{j=1}^{\mathcal{N}-1}(\Delta_{1}c_{B,j}^{\dag}c_{A,j}^{\dag}+\Delta_{2}c_{A,j+1}^{\dag}c_{B,j}^{\dag}+\mathrm{h.c.})\notag\\
&-&\mu\sum_{j=1}^{\mathcal{N}}(c_{A,j}^{\dag}c_{A,j}+c_{B,j}^{\dag}c_{B,j}).\label{1}
\end{eqnarray}
$c_{A,j}(c_{A,j}^{\dag})$ and $c_{B,j}(c_{B,j}^{\dag})$ are annihilation (creation) operators of a fermion at the site $j$, $\mu$ is the chemical potential. $t_{1(2)}$ and $\Delta_{1(2)}$ are the hopping amplitude and SP strength respectively, with the forms of $t_{1(2)}=-t(1\pm\eta)$ and $\Delta_{1(2)}=\Delta(1\pm\eta)$, where $t$ denotes the transfer integral, and $\Delta$ is the SP gap. The hopping and pairing terms are related by the dimerization parameter $\eta$ with the spatial differences $(1\pm \eta)$, and it is set to $|\eta|<1$ to ensure positive transfer integral. In addition, $\cal N$ denotes the length of the dimerized Kitaev chain.

\par
$H_{I}$ is the Hamiltonian of the intercell hopping and SP terms, which can be expressed as
\begin{eqnarray}
H_{I}&=&\sum_{j=1}^{\mathcal{N}-1}(\Delta_{3}c_{B,j+1}^{\dag}c_{A,j}^{\dag}+\mathrm{h.c.})\notag\\
&&+\sum_{j=1}^{\mathcal{N}-1}(t_{3}c_{B,j+1}^{\dag}c_{A,j}+\mathrm{h.c.}).
\end{eqnarray}
Here the intercell hopping amplitude and the SP strength are expressed as $t_{3}$ and $\Delta_{3}$, respectively. It is not difficult to find that due to these two interactions, our structure can be equivalent to an effective ladder lattice, as shown in Fig.1(b). Accordingly, its physics properties will be different from the normal dimerized systems.
\par
Following Eqs.(1)-(3), we can analyze the leading band structure of our system.
In the momentum space, the matrix of the system Hamiltonian can be written in the Bogoliubov-de Gennes form, i.e.,
\begin{eqnarray}
H&=&\sum_{k}C_{k}^{\dag}H(\textbf{k})C_{k},
\end{eqnarray}
where $C_{k}^{\dag}=(c_{k,A}^{\dag},c_{k,B}^{\dag},c_{-k,A},c_{-k,B})$, and $H(\textbf{k})$ can be given as
\begin{small}
\begin{eqnarray}
{H(\textbf{k})}=\left[\begin{array}{cccc}
  -\mu & z & 0 & \omega  \\
  z^{*} & -\mu & -\omega^{*} & 0  \\
  0 & -\omega & \mu & -z  \\
  \omega^{*} & 0 & -z^{*} & \mu \\
\end{array}\right],\notag
\end{eqnarray}
\end{small}
with $z(k)=t_{1}+t_{2}e^{-ik}+t_{3}e^{ik}$ and
$\omega(k)=-\Delta_{1}+\Delta_{2}e^{-ik}-\Delta_{3}e^{ik}$. The lattice constant has been assumed to be one. Diagonalizing $H(\textbf{k})$, its eigenvalues can be written as
\begin{eqnarray}
E^{2}=\mu^{2}+|z|^{2}+|\omega|^{2}\pm2\sqrt{\mu^{2}|z|^{2}+(4t\Delta\eta)^{2}}.
\end{eqnarray}
It can be found that when $k=0$,
$E(0)=\pm[(2t-t_3)\pm\sqrt{\mu^{2}+(\Delta_{3}+2\Delta\eta)^{2}}]$,
and the gap closes at
$\mu^{2}=(2t-t_{3})^{2}-(\Delta_{3}+2\Delta\eta)^{2}$.
While for the case of $k=\pi/a$, there will be
$
E(\pi/a)=\pm[(2t\eta+t_{3})\pm\sqrt{\mu^{2}+(\Delta_{3}-2\Delta)^{2}}]$
with the gap closing at $\mu^{2}=(2t\eta+t_{3})^{2}-(\Delta_{3}-2\Delta)^{2}$.
The gap-closing points correspond to the phase boundaries, which are shown in Fig.2(a)-(b).
\par
Regarding the symmetry of this system, it possesses the time-reversal, particle-hole, and chiral symmetries. And when $\mu=0$, the system also satisfies the sublattice symmetry. And then, according to the symmetry classification, our system belongs to the BDI class since the square of the time-reversal operator and the square of the sublattice symmetry operator equal to $1$. In general, the topological index of the one dimensional system $\mathbb{Z}$-index can be defined by two distinct winding numbers that correspond to two symmetries. Firstly, the winding number caused by the sublattice symmetry can be written as
\begin{eqnarray}
N_{1}&=&\frac{1}{4\pi i}\int_{-\pi/a}^{\pi/a}dk{\rm Tr}[C_{1}H_{k}^{-1}\partial_{k}H_{k}]\notag\\
&=&-\sum_{n=1,2}\int_{-\pi/a}^{\pi/a}\frac{dk}{2\pi i}\partial_{k}\ln z_{n}(k),
\end{eqnarray}
with $z_{1}=A_{1}+B_{1}e^{-ik}+D_{1}e^{ik}$ and $z_{2}=A_{2}+B_{2}e^{-ik}+D_{2}e^{ik}$.
$C_{1}=\tau_{0}\otimes\sigma_{z}$ is the sublattice symmetry operator, $A_{1}=t_{1}-\Delta_{1}$, $B_{1}=t_{2}+\Delta_{2}$, $D_{1}=t_{3}-\Delta_{3}$, and $A_{2}=-t_{1}-\Delta_{1}$, $B_{2}=-t_{2}+\Delta_{2}$, $D_{2}=-t_{3}-\Delta_{3}$.
On the other hand, the winding number induced by the particle-hole symmetry of the superconductor reads
\begin{eqnarray}
N_{2}&=&\frac{1}{4\pi i}\int_{-\pi/a}^{\pi/a}dk{\rm Tr}[C_{2}H_{k}^{-1}\partial_{k}H_{k}]\notag\\
&=&-\sum_{n=1,2}\int_{-\pi/a}^{\pi/a}\frac{dk}{2\pi i}\partial_{k}\ln Z(k),
\end{eqnarray}
where $Z(k)=-\mu^{2}+(z-\omega)(z^{*}+\omega^{*})$ with $C_{2}=\tau_{0}\otimes\sigma_{x}$ being the particle-hole symmetry operator.

\section{Numerical results and discussions \label{result2}}
In this section, we proceed to discuss the topological properties of the dimerized Kitaev chain with the intercell hopping and SP terms on the basis of the theory in Sec. II. Our purpose is to clarify the roles of these two interactions in changing the topological property of this system. For calculation, we take $t=1.0$ as the energy unit.
\subsection{$t_{3}\neq0$ but $\Delta_{3}=0$}
\begin{figure}[htb]
\centering \scalebox{0.17}{\includegraphics{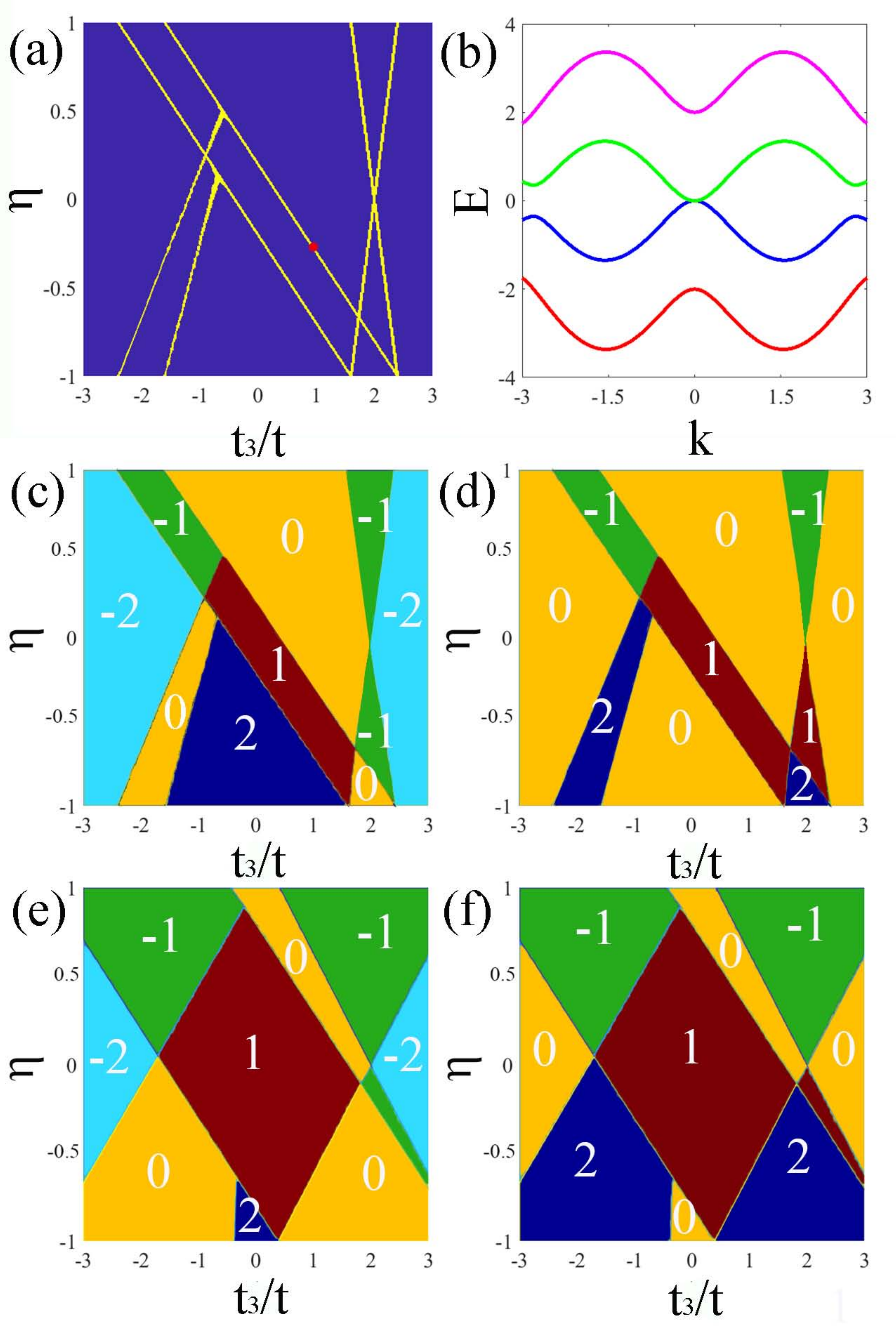}} \caption{(a) Energy gap-closing points with the increase of the intercell hopping amplitude $t_{3}$ and $\eta$ in the case of $\Delta=0.2$. (b) The energy dispersions as a function of momentum $k$ in the case of $\Delta=0.2$, $t_{3}=1.0$ and $\eta=-0.2$. (c), (e) Topological phase diagrams with respect to $N_{1}$ for $\Delta=0.2$ and $0.8$, respectively. (d), (f) Topological phase diagrams with respect to $N_{2}$ for $\Delta=0.2$ and $0.8$, respectively. Other parameters are taken to be $\mu=0$, $\Delta_{3}=0$.
\label{fig2}}
\end{figure}
\par
\begin{figure*}[htb]
\centering \scalebox{0.22}{\includegraphics{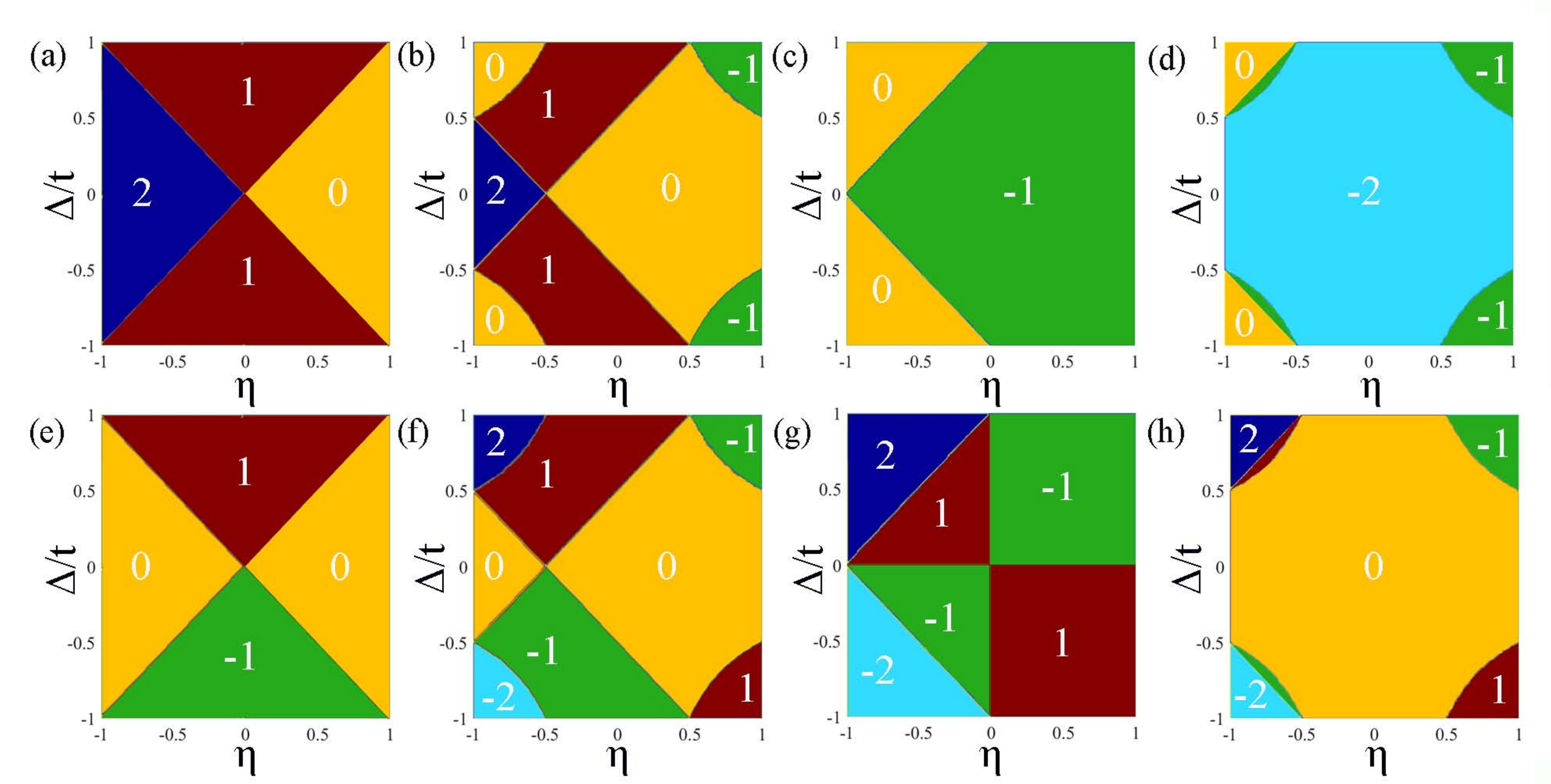}} \caption{Topological phase diagrams for $N_{1}$ (top panel) and $N_{2}$ (bottom panel) caused by the increase of $\eta$ and $\Delta$, the parameters are taken to be (a), (e) $t_{3}=0$; (b), (f) $t_{3}=1.0$; (c), (g) $t_{3}=2.0$; and (d), (h) $t_{3}=3.0$. Other parameters are $\mu=0$ and $\Delta_{3}=0$.
\label{fig2}}
\end{figure*}
To begin with, we would like to consider the presence of intercell hopping term to investigate the topological property of the dimerized Kitaev chain. Firstly, we take the case of $\Delta=0.2$ with $\mu=0$ and calculate the dependence of the gap-closing points on the intercell hopping parameters $t_{3}$ and $\eta$ by taking $\Delta_{3}=0$. The numerical result is shown in Fig.2(a). And for comparison, we choose the case of $t_{3}=1.0$ and $\eta=-0.2$ to present the appearance of the gap-closing point in Fig.2(b). It can be found that in such a case, the band gap is closed at the point $k=0$. Thus, the results in Fig.2(a) suggest that both $t_3$ and $\eta$ make nontrivial contribution to the appearance of the gap-closing points, which are the signs of the topological phase transition. In order to investigate the concrete topological phases, we plot the diagrams of the winding numbers $N_1$ and $N_2$ as functions of $t_3$ and $\eta$ [see Fig.2(c)-(d)]. Since the winding numbers depend on the choice of the global phase, the signs of them are meaningless, and we can classify the topological phases by the absolute values of the winding numbers. It can be found that in the region of $|N_1|=1$, $|N_2|$ is also equal to 1. And when one equals to $2$, the other becomes equal to zero. Hence in Fig.2(c)-(d), there exists three topological nontrivial phases: (i) SSH-like phase where $|N_{1}|=2$ and $|N_{2}|=0$; (ii) Kitaev-like phase with $|N_{1}|=1$ and $|N_{2}|=1$; (iii) degenerated Kitaev-like phase with $|N_{1}|=0$ and $|N_{2}|=2$. Nevertheless, the topological trivial phase comes into being where $|N_{1}|=|N_{2}|=0$. Note that compared with the general dimerized Kitaev chain, the intercell hopping terms produce a new phase, i.e., the degenerated Kitaev-like phase where $|N_{1}|=0$ and $|N_{2}|=2$. It means that in this phase, there are two degenerated MZMs at each end of the chain. Fig.2(c)-(d) also show that the topological phases are asymmetric about $\pm\eta$ and $\pm t_{3}$, and the degenerated Kitaev-like phase mainly appears when $\eta$ is negative and $t_{3}$ near $\pm2$. In the case of $t_{3}\approx-2.0$, the degenerated Kitaev-like phase appears during $\eta<0.2$, while around the position $t_{3}\approx2.0$, the Kitaev-like phase comes up if $\eta<-0.7$. Surely, the region of the Kitaev-like topological phase is smaller than that of the SSH-like topological phase due to the smaller SP strength. As $\Delta$ increases to $0.8$, the results in Fig.2(e)-(f) show that the regions of the Kitaev-like and degenerated Kitaev-like phases are expanded  efficiently. It can be clearly seen that the degenerated Kitaev-like phase appears in the wider regions where $t_{3}<-0.2$ and $\eta<0.1$ or $t_{3}>0.2$ and $\eta<-0.1$, and it is almost symmetric with the regions of the Kitaev-like phase about $\eta=0$. One can therefore get the first sight about the interplay between $t_3$ and $\Delta$ in driving the topological phase transition. What is more, the distribution of $N_{1}$ obeys the relationship
\begin{eqnarray}
N_{1}&=&\Theta D_{1}+\Theta D_{2}-\Theta E_{1}-\Theta E_{2}\notag\\
&&+\Theta(E_{1}F_{1})+\Theta(E_{2}F_{2})-2,
\end{eqnarray}
where $\Theta(x)$ denotes the heaviside step function, and $D_{1}=2(t+\Delta\eta)-t_{3}+\Delta_{3}$, $D_{2}=2(t-\Delta\eta)+t_{3}-\Delta_{3}$, $E_{1}=(\Delta-t)(1-\eta)-t_{3}+\Delta_{3}$, $E_{2}=-(\Delta+t)(1-\eta)-t_{3}-\Delta_{3}$, $F_{1}=-2(t\eta+\Delta)+t_{3}-\Delta_{3}$, $F_{2}=2(t\eta-\Delta)+t_{3}+\Delta_{3}$. This expression implies the complicated interplay among the parameters in driving the topological phase transitions.
\par

\par
Inspired by the results in Fig.2, we next pay attention to the phase diagram contributed by the interplay between $\eta$ and $\Delta$, as shown in Fig.3. In this figure, the intercell hopping amplitude is taken to be $t_{3}=0$, $1.0$, $2.0$, and $3.0$, respectively. The result of $N_1$ is exhibited in Fig.3(a)-(d) and that of $N_{2}$ is presented in Fig.3 (e)-(h) correspondingly. We see that each phase is symmetric about the line of $\Delta=0$, and the degenerate Kitaev-like phase appears only when $\eta<0$, which is accordant with the results in Fig.2. Now let us focus on the topological phase transition caused by the intercell hopping terms. Firstly when $t_{3}=0$, three phases are allowed to exist with the change of $\Delta$ and $\eta$, i.e., the Kitaev-like phase, the SSH-like topological phase, and the topological trivial phase. As $t_{3}$ increases to $1.0$, the degenerated Kitaev-like phase appears
in the position $\eta<-0.5$ with $|\Delta|>0.5$. Meanwhile, in the regions of $\eta>0.5$, the Kitaev-like phase begins to come up. All these changes are accompanied by the narrowness of the regions of SSH-like and topologically trivial phases. It is worth noticing that when $t_{3}=2.0$, the SSH-like and topologically trivial phases disappear, whereas the ranges of the other two phases reach their extremum. Excess this critical case, the alternative
phenomenon will take place. For instance, when $t_{3}=3.0$, the SSH-like phase
becomes dominant in the whole energy region, followed by the suppression of the topological
trivial phase. So far, we know that due to the existance of the intercell hopping term, the
topological properties of the dimerized Kitaev chain can be enriched in a substantial way.
\par
\begin{figure}[htb]
\centering \scalebox{0.19}{\includegraphics{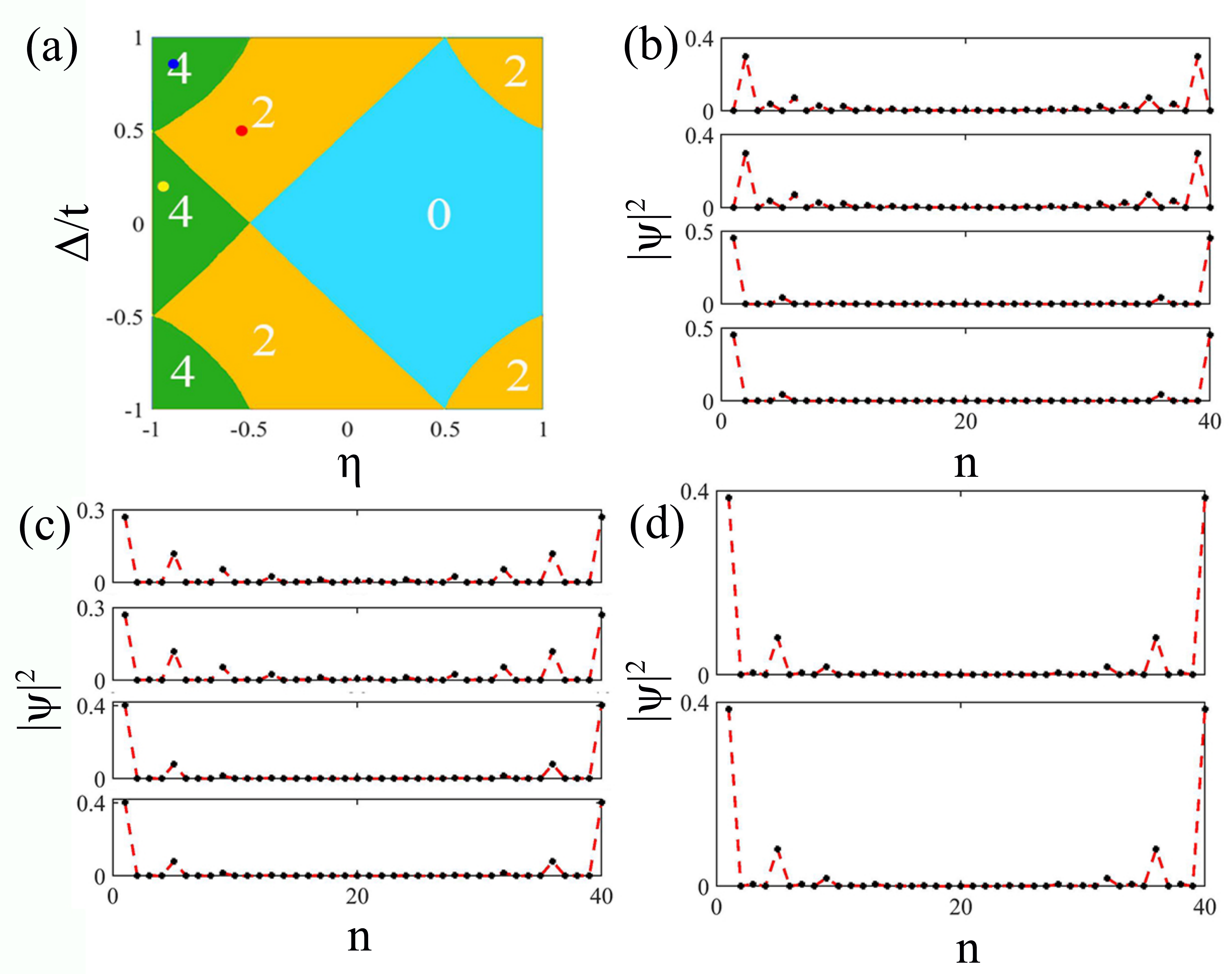}} \caption{(a) The number of zero-energy modes as a function of $\eta$ and $\Delta$. The distributions of zero-energy modes for (b) $\eta=-0.8, \Delta=0.8$, (c) $\eta=-0.9, \Delta=0.2$, and (d) $\eta=-0.5, \Delta=0.5$.
Other parameters are $t_{3}=1.0$, $\mu=0$, $\Delta_{3}=0$, ${\cal N}=40$. Three dots with different colors are plotted, exactly corresponding to three kinds of topological phases. \label{fig4}}
\end{figure}
In order to clarify the relationship between the winding numbers $N_{1(2)}$ and the zero-energy mode number, we exhibit the distributions of zero modes for the Kitaev-like phase, the SSH-like phase, and the degenerated Kitaev-like phase, manifested as the dependence of the zero-energy mode number on the winding numbers. It does show that in the three topological phase regions, the zero-energy mode number can reach 4 and 2, respectively. For further presenting the properties of these zero modes, we plot their distributions in the real space. The corresponding results are shown in Fig.4(b) $\eta=-0.8$ and $\Delta=0.8$, (c) $\eta=-0.9$ and $\Delta=0.2$, and (d) $\eta=-0.5$ with $\Delta=0.5$ and $t_{3}=1.0$, and the length number of the chain is set to ${\cal N}=40$. The different zero modes correspond to the yellow, red, and blue dots marked in Fig.4(a) for convenience. In Fig.4(b), we notice that there are fourfold degenerated MZMs, corresponding to the degenerated Kitaev-like phase. And two of the MZMs are localized at the end cites of the chain, but the other two appear at the subend sites. Next in Fig.4(c), it shows that fourfold degenerated fermionic zero modes (FZMs) appear, corresponding to the SSH-like topological phase, where four FZMs are localized at the end cites of the chain. As for the Kitaev-like phase, we can see that there are only twofold degenerated MZMs localized at the end cites of the chain, as shown in Fig.4(d).
\par
\begin{figure}[htb]
\centering \scalebox{0.22}{\includegraphics{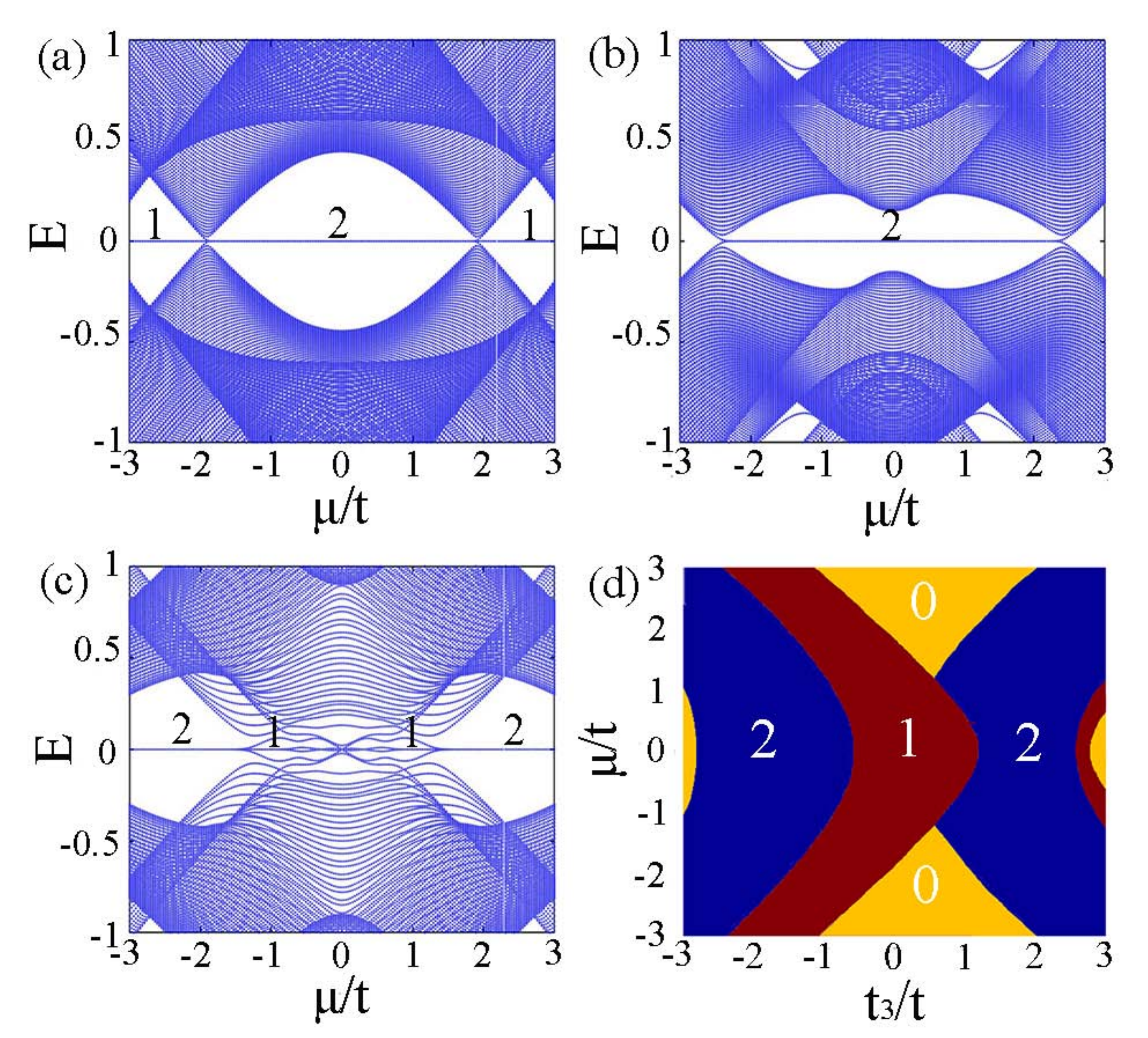}} \caption{Energy spectrum in real space for (a) $t_{3}=-1.5$, (b) $t_{3}=1.5$ and (c) $t_{3}=3.0$. (d) Topological phase diagrams with respect to $N_{2}$ caused by the increase of $t_{3}$ and $\mu$. The parameters are taken to be $\Delta_{3}=0$, $\Delta=0.8$, $\eta=-0.5$ and ${\cal N}=120$.
\label{fig5}}
\end{figure}
Next, we would like to consider the influence induced by the nonzero chemical potential, i.e., $\mu\neq0$. The numerical results are shown in Fig.5. Because the nonzero chemical potential destroys the sublattice symmetry, the SSH-like topological phase will be suppressed, thus we only need to calculate the value of $N_{2}$. Firstly, we plot the energy spectrum with respect to the chemical potential $\mu$, as shown in Fig.5(a)-(c), where the values of intercell hopping amplitude $t_{3}$ are set to be $-1.5$, $1.5$, $3.0$, respectively. It is found that when $t_{3}=-1.5$, the zero-energy modes are twofold-degenerated in the range of $2.0<|\mu|<3.0$, and fourfold-degenerated in the range of $-2<\mu<2$, where $\mu=\pm2$ are the topological phase transition points, as shown in Fig.5(a). Next if $t_{3}=1.5$, Fig.5(b) shows that there are fourfold degenerated zero modes appearing in the range of $-2.5<\mu<2.5$, while none exists elsewhere. As $t_{3}$ increases to $3.0$, it shows that the three phases coexist, namely, the zero-energy states are fourfold-degenerated in the range of $1.2<|\mu|<3$, and twofold-degenerated in the range of $0.8<|\mu|<1.2$, and no zero-energy mode exists in the range of $-0.8<\mu<0.8$, as shown in Fig.5(c). More results can be observed in the topological phase diagram of Fig.5(d). It can be found that the intercell hopping terms play an important role in modulating the topological phase transition and the size of the energy gap.

\par
\begin{figure}[t]
\centering \scalebox{0.19}{\includegraphics{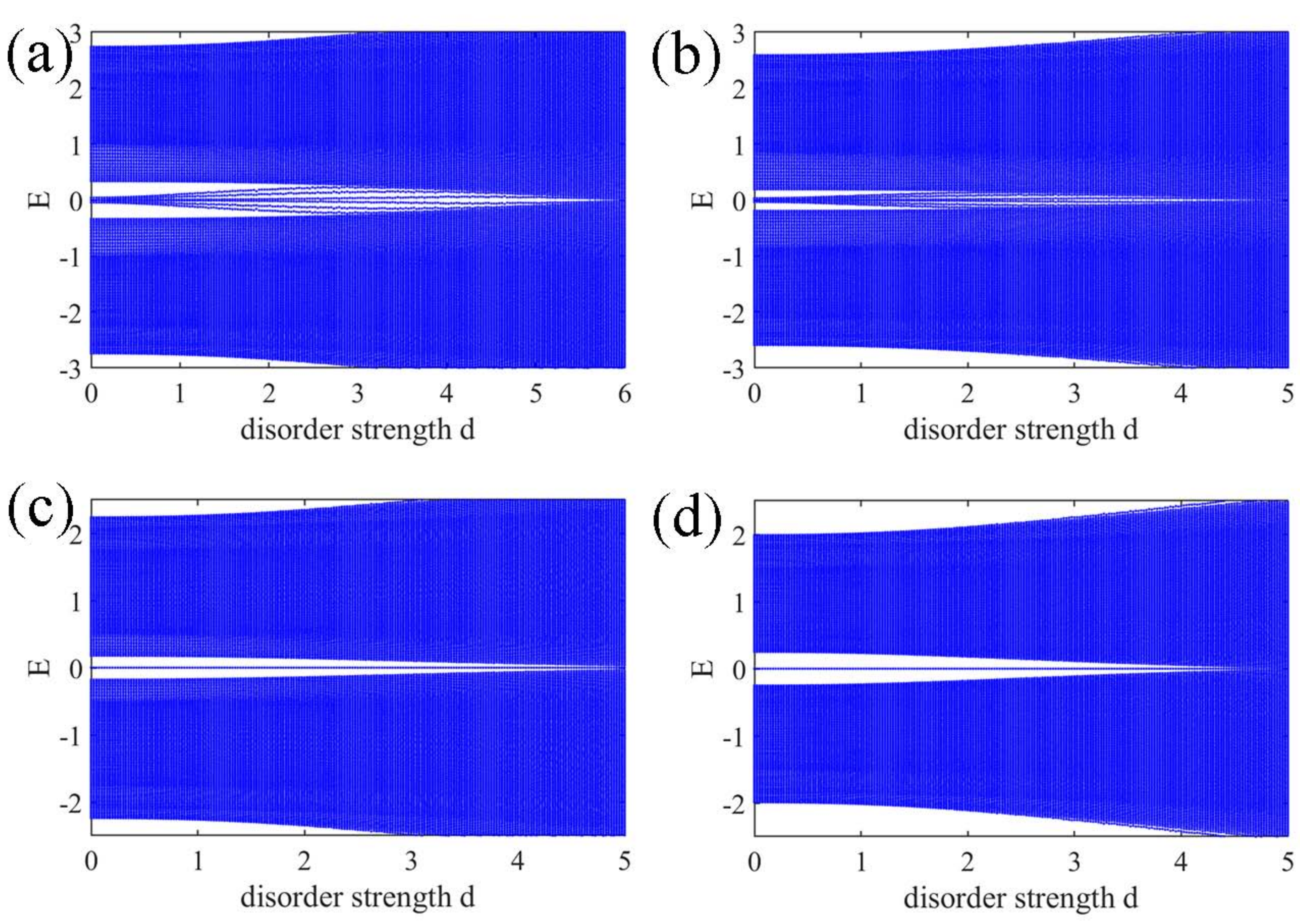}} \caption{Band structure influenced by the increasing of disorder strength $d$ for (a) $t_{3}=-3$, (b) $t_{3}=-2.7$, (c) $t_{3}=-2$, (d) $t_{3}=-1.5$. The other parameters are $\mu=0$, $\eta=-0.3$, $\Delta=0.8$, $\Delta_{3}=0$, and ${\cal N}=120$. \label{fig6}}
\end{figure}
\par
Since the ideal systems are difficult to realize in experiment, in Fig.6 we would like to study the disorder effect\cite{disorder1,disorder2,disorder3} on our system to see whether the zero-energy modes are robust to disorder or not. The disorder is introduced by replacing the chemical potential $\mu$ with the disorder term $\mu_{j}=\mu+d\omega_{j}$, where $d$ is the disorder strength, and $\omega_{j}$ is the disorder distribution in the range of $[-0.5,0.5]$.
Fig.6(a)-(d) correspond to the energy spectrum influenced by the increase of disorder strength $d$ for $t_{3}=-3$, $-2.5$, $-2$ and $-1.5$, respectively. According to Fig.2(e)-(f), we know that Fig.6 (a)-(b) are related to the SSH-like phase, and Fig.6(c)-(b) describe the degenerated Kitaev-like phase. This exactly means that they show the disorder effect on the FZMs and the MZMs, respectively. We observe in Fig.6(a)-(b) that with the increase of the disorder strength, the energies of FZMs gradually integrate into the bulk band and meanwhile the band of the bulk states is widened. All these results cause the FZMs to disappear. Furthermore, the FZMs are affected by the intercell hoppings, that is, the FZMs are broken by disorder more easily when $t_{3}$ increases. On the other hand, for the MZMs, in Fig.6(c)-(d) it shows that the energies of the MZMs are independent of the increase of the disorder strength and the intercell hoppings. The disorder effect only widens the band of the bulk states.  For this reason, when the disorder is strong enough, the MZMs can also be destroyed. Note, however, that the disorder presents different effects for the situations of FZMs and MZMs, due to the direct disappearance of the FZMs. For instance, in the case of $t_3=-1.5$ with $d\le2.0$, the band structure of our system is almost robust, accompanied by the clear signature of the MZMs.

\subsection{$t_{3}\neq0$ and $\Delta_{3}\neq0$}
\begin{figure}[htb]
\centering \scalebox{0.17}{\includegraphics{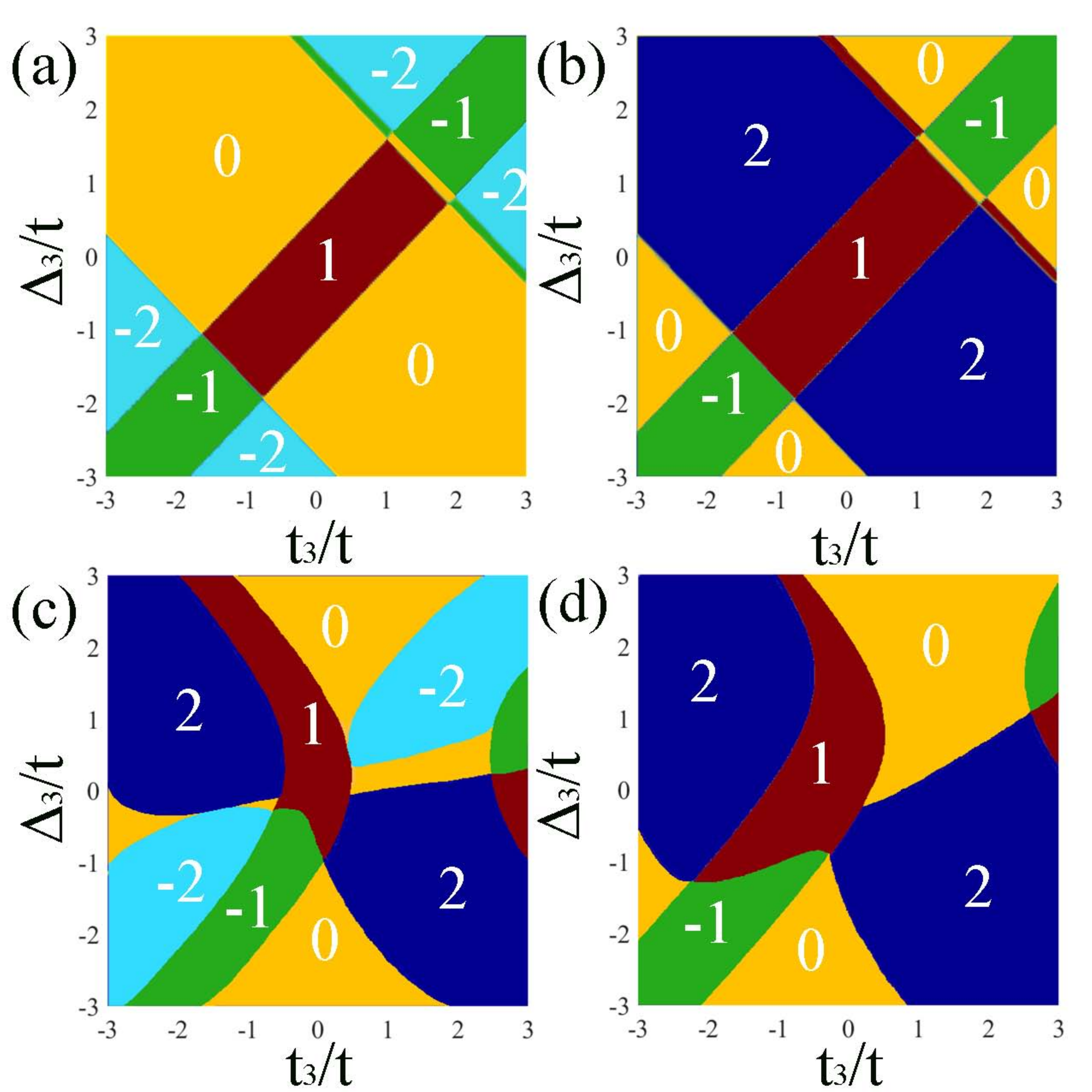}} \caption{(a)-(b) Topological phase diagrams with respect to $N_{1}$ and $N_{2}$ caused by the increase of $t_{3}$ and $\Delta_{3}$ in the case of $\mu=0$, $\eta=-0.5$, and $\Delta=0.8$. (c)-(d) Topological phase diagrams of $N_{2}$ for $\Delta=0.2$ and $0.8$, respectively. The other parameters are taken to be $\mu=-1.5$ and $\eta=-0.5$. \label{structure}}
\end{figure}
In this part, we would like to introduce the intercell SPs, and investigate the role it plays in modifying the topological properties of the present system.
\par
In Fig.7 we introduce the intercell SPs and investigate the topological phase diagrams caused by the interplay between the intercell hopping amplitude $t_{3}$ and the intercell SP strength $\Delta_{3}$ to clarify the roles of them in the topological phases. Fig.7(a)-(b) show the phase diagrams of $N_{1}$ and $N_{2}$ in the case of $\mu=0$, with the parameters being $\eta=-0.5$ and $\Delta=0.8$. We see that four topological phases appear, and the region of the topological trivial phase is negligibly small.
And also, the respective phase regions are almost symmetric about the line of $\Delta_{3}=t_{3}$, which means the identical contributions of the intercell hopping and SP terms to the topological phase transition. That is to say, the intercell SPs can also lead to the degenerated Kitaev-like phase and similar topological phase transition in comparison with the intercell hopping terms when $\mu=0$. Next, Fig.7(c)-(d) exhibit the phase diagrams of $N_{2}$ for $\Delta=0.2$ and $0.8$, respectively, with the parameters $\mu=-1.5$ and $\eta=-0.5$. It can be seen that this two phase diagrams do not obey any symmetries, and three topological phases appear in each phase diagram, i.e., the Kitaev-like and degenerated Kitaev-like phase, and the topological trivial phase. For the case of $\Delta=0.2$, it can be seen that when $\Delta_{3}=0$, changing $t_{3}$ in the range of $-3<t_{3}<-0.5$ induces the degenerated Kitaev-like phase. With the increase of $|\Delta_{3}|$, such a range increases accordingly. As $\Delta=0.8$, the situation becomes more complicated, and the range of $t_{3}$ does not vary monotonously with the increment of $\Delta_{3}$. However, it decreases with the increase of $|\Delta_{3}|$ except in the range of $-1<\Delta_{3}<0$. The above phenomena indicate that $\Delta_{3}$ does play an important role in regulating the degenerated Kitaev-like phase, and it mainly expands the region of this phase in the case of small $\Delta$.
\par
\begin{figure}[t]
\centering \scalebox{0.17}{\includegraphics{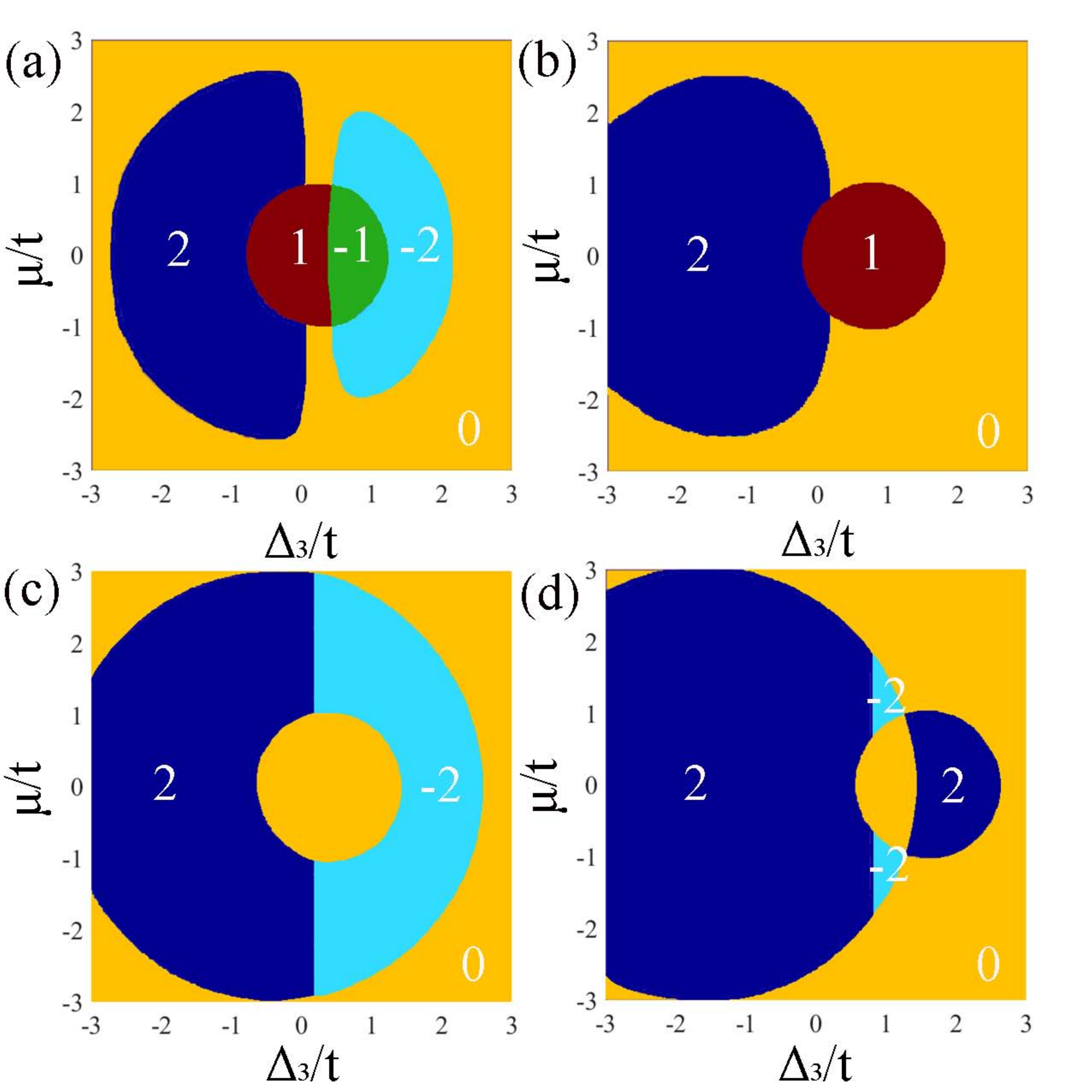}} \caption{Topological phase diagrams with respect to $N_{2}$ caused by the increase of $\Delta_{3}$ and $\mu$ for $t_{3}=1$, the parameters are taken to be (a) $\Delta=0.2$, $\eta=-0.5$; (b) $\Delta=0.8$, $\eta=-0.5$; (c) $\Delta=0.2$, $\eta=-1$; d) $\Delta=0.8$, $\eta=-1$. \label{fig7}}
\end{figure}
\begin{figure}[t]
\centering \scalebox{0.21}{\includegraphics{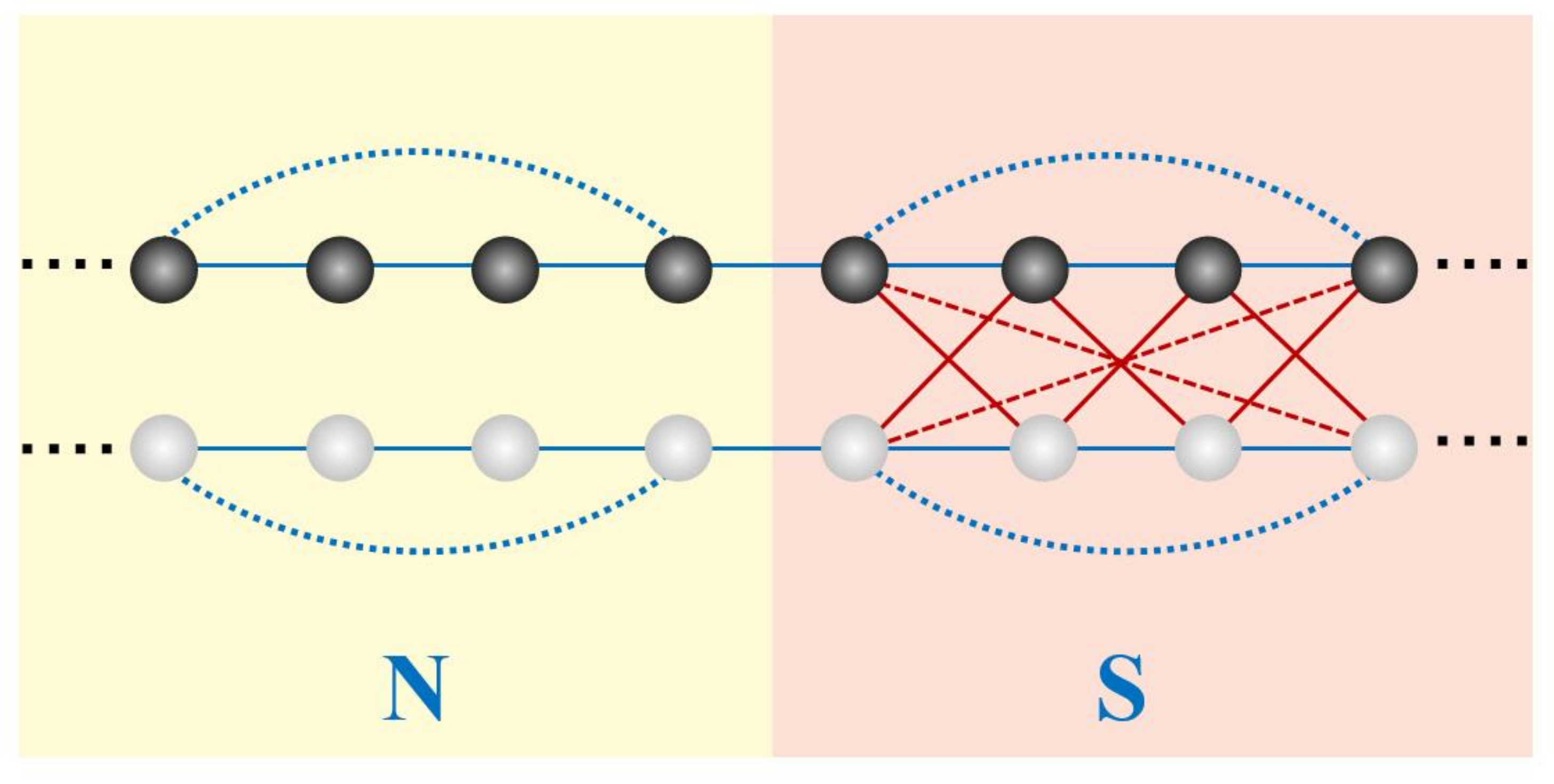}} \caption{Illustration of the N-TS junction of oue considered structure in the Nambu representation. The black balls denote the electronic states, and the white balls are the hole counterpart. \label{fig8}}
\end{figure}
\par
Subsequently, we calculate the topological phase diagrams of $N_{2}$ as a function of the intercell SP strength $\Delta_{3}$ and chemical potential $\mu$ for several values of $\Delta$ and $\eta$ under the condition $t_{3}=1.0$, as shown in Fig.8. Firstly in Fig.8(a)-(b), we present the phase diagrams for $\eta=0.5$, $\Delta=0.2$ and $0.8$, respectively. It is clearly shown that in these two figures, three phases appear, i.e., the Kitaev-like phase, the degenerated Kitaev-like phase, and the topological trivial phase. When $\Delta=0.2$, the system is located in the Kitaev-like phase or topological trivial phase for $\Delta_{3}=0$, and can enter the degenerated Kitaev-like phase by increasing $|\Delta|$. In the case of $\Delta=0.8$, the system mainly exhibits the degenerated Kitaev-like phase for negative $\Delta_{3}$. Next in Fig.8(c)-(d), the parameters are set to $\eta=-1$, $\Delta=0.2$ and $0.8$, respectively. Compared with Fig.8(a)-(b), the Kitaev-like phase disappears. As shown in Fig.8(c), the degenerated Kitaev-like phase appears in the form of a circle with a radius of 3.0. If $\Delta$ increases to $0.8$, Fig.8(d) shows that the region of the degenerated Kitaev-like phase moves towards the negative-$\Delta_{3}$ direction, and the region of the topological trivial phase shifts to the positive-$\Delta_{3}$ direction. In totally, the region of the degenerated Kitaev-like phase decreases but the topological trivial phase is enhanced. Thus $\Delta_{3}$ can effectively modulate the correspondence of $\mu$ to the degenerated Kitaev-like phase. Moreover, the results of Fig.7 and Fig.8 indicate that the intercell SPs can magnify the region of the twofold degenerated Kitaev-like phase, in the presence of the appropriate parameter ranges.
\par
\begin{figure}[t]
\centering \scalebox{0.17}{\includegraphics{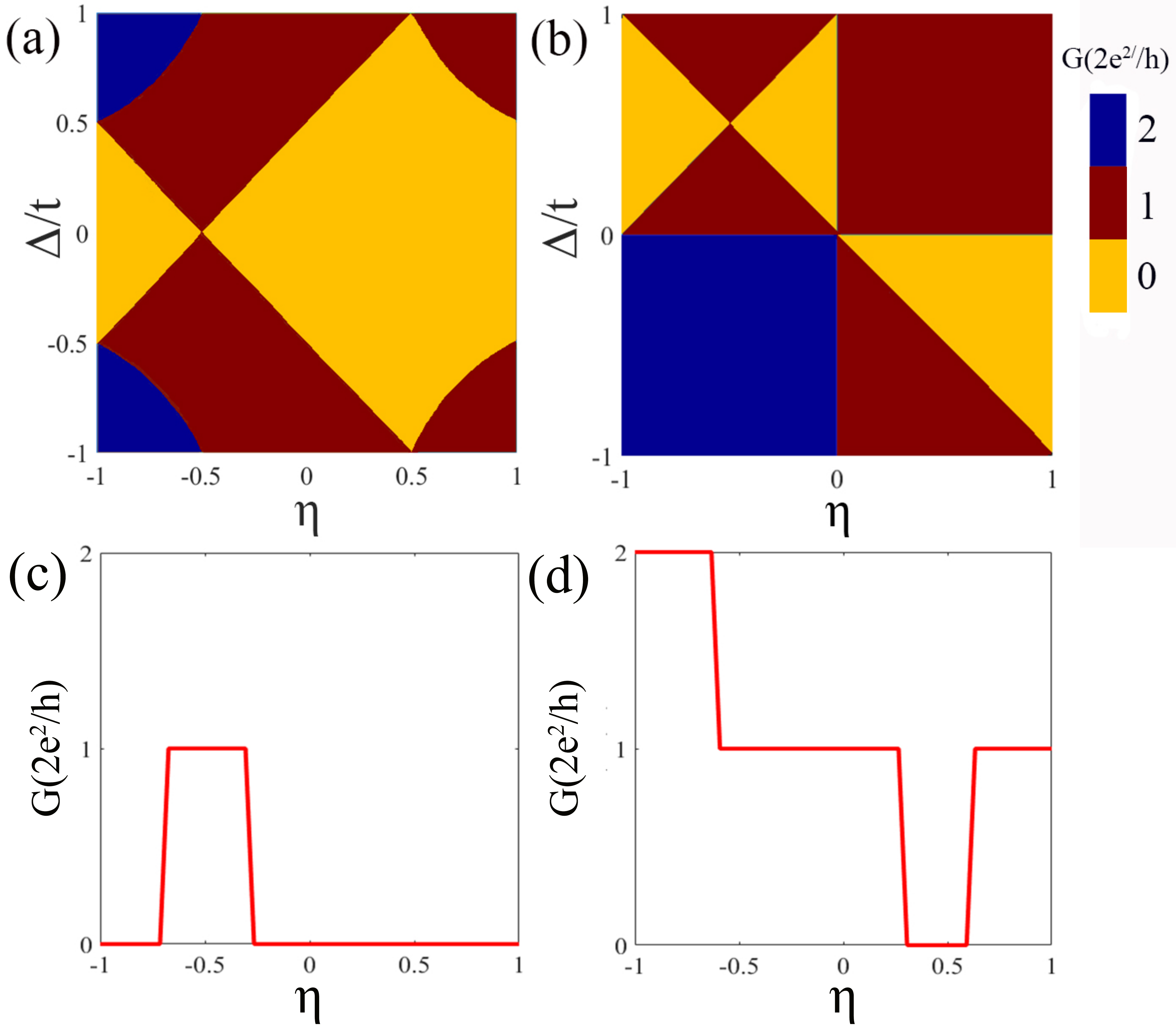}} \caption{The zero-bias Andreev-reflection conductance caused by the increase of $\eta$ and $\Delta$ in the cases of (a) $\Delta_{3}=0$ and (b) $\Delta_{3}=1$. The zero-bias differential conductance as a function of $\eta$ for  (c) $\Delta=0.2$, $\Delta_{3}=0$ and (d) $\Delta=0.8$, $\Delta_{3}=0$. The other parameters are $\mu=0$ and $t_{3}=1$. \label{fig8}}
\end{figure}

\subsection{Andreev-reflection conductances}
\par
In order to examine the phase results of our system, we design a N-TS junction composed by two semi-infinite dimerized chains with or without the SP terms. The illustration in the Nambu representation is shown in Fig.9. With the help of the recursive Green's function method\cite{recursive1,recursive2,recursive3}, we calculate the Andreev-reflection conductance spectra in Fig.10. The Hamiltonian of the superconducting part can be described by Eq.(1), and the normal part has the same Hamiltonian with $\Delta=\Delta_{3}=0$. Furthermore, the hopping amplitude between the two parts is set to $t$. At the zero-temperature limit, the Andreev-reflection conductance can be given as
\begin{equation}
{\cal G}={2e^2\over h}T_{eh}(\omega=eV)
\end{equation}
with $T_{eh}={\mathrm{Tr}}[\Gamma_e G \Gamma_h G^\dag]$. $G$ is
the retarded Green's function, and it is formally given by the Dyson equation
\begin{equation}
G(\omega)=[\omega+i0^+-H-\Sigma]^{-1},
\end{equation}
where the retarded self-energy $\Sigma=\Sigma_e+\Sigma_h$ arises from the coupling between the two parts.
$\Gamma_\alpha=i[\Sigma_\alpha-\Sigma_\alpha^\dag]$ with $\alpha=e,h$ is the matrix of the electronic(hole)-state component of the coupling strength.
\par
It is known that the key step to solve the retarded Green's function is the solution of $\Sigma_\alpha$. We would like to point out that
\begin{equation}
\Sigma_\alpha=H^\alpha_{\mathrm{TS-N}}g_{0\alpha}H^\alpha_{\mathrm{N-TS}},
\end{equation}
in which $H^\alpha_{\mathrm{TS-N}}$ is the $\alpha$-component of the coupling between the two parts. $g_{0\alpha}$ is surface Green's function of the semi-infinite dimerized chain, which can be solved as follows. Namely,
\begin{eqnarray}
g_{0\alpha}=[\omega+i0^+-H_{0\alpha}-H_{01,\alpha}g_{1\alpha}H_{10,\alpha}]^{-1},
\end{eqnarray}
where $g_{1\alpha}$ is the Green's function of the $\alpha$-component of the sub-surface layer. As for $g_{1\alpha}$, it obeys the similar equation
\begin{eqnarray}
g_{j\alpha}=[\omega+i0^+-H_{j\alpha}-H_{j,j+1,\alpha}g_{j+1,\alpha}H_{j+1,j,\alpha}]^{-1}.\notag
\end{eqnarray}
For the periodic chain, one can solve the surface Green's function iteratively, by supposing $g_{j\alpha}=g_{j+1,\alpha}$.

Through the above equations, we plot the spectra of the zero-bias Andreev conductance in the case of sublattice symmetry, as shown in Fig.10(a)-(b), respectively. Firstly in Fig.10(a), it can be observed that the conductance value is consistent with the absolute value of $N_{2}$, whereas it is irrelevant to $N_{1}$. This does indicate that only the MZMs contribute to the zero-bias Andreev conductance. Furthermore, the magnitude of $\cal G$ can reach $4e^{2}/h$, which is caused by the twofold-degenerated MZMs emerging in the degenerated Kitaev-like phase. Next in Fig.10(b), when $\Delta_{3}=t_{3}=1.0$, the distribution of the zero-bias Andreev conductance changes a lot, and ${\cal G}=4e^{2}/h$ when both $\Delta$ and $\eta$ are negative. For further presenting the conductance properties, we take $\Delta=0.2$ and $0.8$, respectively, and plot the curves of the zero-bias conductance as a function of $\eta$ in the case of sublattice symmetry, as shown in Fig.10(c)-(d). The other parameters are the same as those in Fig.10(a). One can clearly observe the step-like variations of the zero-bias Andreev conductance, with the increase of $\eta$. Up to now, it can be ascertained that the properties of the MZMs in this system can be directly clarified by measuring the Andreev-conductance spectra.

\section{summary\label{summary}}
To summarize, we have investigated the topological properties of the dimerized Kitaev chain, by considering the long-range intercell hopping and SP terms. As a result, it has been found that these two mechanisms exhibit abundant phase transitions, with the adjustment of the structural parameters. To be concrete, even only when the intercell hopping terms are incorporate, the proportion of topological phases can be modulated, leading to the topological phase transition. The notable result is that one new Kitaev-like phase can appear in this system, which possesses twofold-degenerated Majorana zero-energy modes localized at each end of the dimerized Kitaev chain. This phase tends to be weakly dependent on the weak disorder.
Next the long-range SPs are found to magnify the region of the new Kitaev-like phase under the appropriate structural parameters. We believe that this work can be helpful for understanding the effect of the intercell hopping and SP terms on topological properties of the dimerized Kitaev systems.

\section*{Acknowledgments}
This work was financially supported by the LiaoNing Revitalization Talents Program (Grant No. XLYC1907033), the National Natural Science Foundation of China (Grants No. 11905027 and 11604221), and the Fundamental Research
Funds for the Central Universities (Grants No. N2002005 and N180503020).

\clearpage

\bigskip


\begin{thebibliography}{99}
\bibitem{high energy1} W. Y. Keung, and G. Senjanovi$\acute{c}$, Phys. Rev. Lett. \textbf{50}, 1427 (1983).
\bibitem{high energy2} A. Atre, T. Han, S. Pascoli, and B. Zhang. J. High Energy Phys. \textbf{030}, 2009 (2009).
\bibitem{high energy3} W. Rodejohann, and Inter. J. Mod. Phys. E \textbf{20}, 1833 (2011).
\bibitem{TS1} R. M. Lutchyn, J. D. Sau, and S. Das Sarma, Phys. Rev. Lett. \textbf{105}, 07001 (2010).
\bibitem{TS2} A. Das, Y. Ronen, Y. Most, Y. Oreg, M. Heiblum, and H. Shtrikman, Nat. Phys. \textbf{8}, 887 (2012).
\bibitem{TS3} M. Sato, and S. Fujimoto, Phys. Rev. B \textbf{79}, 094505 (2009).
\bibitem{TS4} J. P. Xu, M. X. Wang, Z. L. Liu, J. F. Ge, X. Yang, C. Liu, Z. A. Xu, D. Guan, C. L. Gao, D. Qian, Y. Liu, Q. H. Wang, F. C. Zhang, Q. K. Xue, and J. F. Jia, Phys. Rev. Lett. \textbf{114}, 017001 (2015).
\bibitem{non-Abelian} G. Moore, and N. Read, Nucl. Phys. B \textbf{360}, 362 (1991).
\bibitem{fault-tolerant} J. Alicea, Rep. Prog. Phys. \textbf{75}, 076501 (2012).
\bibitem{MZMs1} M. Alidoust, M. Willatzen, and A. P. Jauho, Phys. Rev. B \textbf{98}, 085414 (2018).
\bibitem{MZMs2} A. Fornieri, A. M. Whiticar, F. Setiawan, E. Portol$\acute{e}$s, A. C. C. Drachmann, A. Keselman, S. Gronin, C. Thomas, T. Wang, R. Kallaher, $et$ $al$., Nature \textbf{569}, 89 (2019).
\bibitem{MZMs3} A. Nava, R. Giuliano, G. Campagnano, D. Giuliano, Phys. Rev. B \textbf{95}, 155449 (2017).
\bibitem{MZMs4} N. Wu, and W. L. You, Phys. Rev. B \textbf{100}, 085130 (2019).
\bibitem{1D-SC1} A. Y. Kitaev, Ann. Phys. \textbf{303}, 2-30 (2003).
\bibitem{1D-SC2} C. Nayak, S. H. Simon, A. Stern, M. Freedman, S. Das Sarma, Rev. Mod. Phys. \textbf{80}, 1083 (2008).
\bibitem{vortex1} N. Read, and D. Green, Phys. Rev. B \textbf{61}, 10267 (2000).
\bibitem{vortex2} D. A. Ivanov, Phys. Rev. Lett. \textbf{86}, 268 (2001).
\bibitem{vortex3} A. Stern, F. von Oppen, and E. Mariani, Phys. Rev. B \textbf{70}, 205338 (2004).
\bibitem{vortex4} M. Stone, and S. B. Chung, Phys. Rev. B \textbf{73}, 014505 (2006).
\bibitem{Oreg} Y. Oreg, G. Refael, F. von Oppen, Phys. Rev. Lett. \textbf{105}, 177002 (2010).
\bibitem{Pientka} F. Pientka, A. Keselman, E. Berg, A. Stern, B. I. Halperin, Phys. Rev. X \textbf{7}, 021032 (2017).
\bibitem{Delft} V. Mourik, K. Zuo, S. M. Frolov, S. R. Plissard, E. P. A. M. Bakkers, Science \textbf{336}, 1003 (2012).
\bibitem{2D-TS1} L. Fu, and C. L. Kane, Phys. Rev. Lett. \textbf{102}, 216403 (2009).
\bibitem{2D-TS2} A. R. Akhmerov, J. Nilsson, and C. W. J. Beenakker, Phys. Rev. Lett. \textbf{102}, 216404 (102).
\bibitem{D1} J. C. Budich, E. Ardonne, Phys. Rev. B \textbf{88}, 075419 (2013).
\bibitem{D2} T. Fukui, T. Fujiwara, Phys. Rev. B \textbf{82}, 184536 (2010).

\bibitem{Zhang2} F. Zhang, C. L. Kane, and E. J. Mele, Phys. Rev. Lett. \textbf{111}, 056403 (2013).
\bibitem{Fuliang2} A. Keselman, L. Fu, A. Stern, and E. Berg, Phys. Rev. Lett. \textbf{111}, 116402 (2013).
\bibitem{Deng} S. Deng, L. Viola, and G. Ortiz, Phys. Rev. Lett. \textbf{108}, 036803 (2012).
\bibitem{Nakosai} S. Nakosai, Y. Tanaka, and N. Nagaosa, Phys. Rev. Lett. \textbf{108}, 147003 (2012).
\bibitem{Wong} C. L. M. Wong and K. T. Law, Phys. Rev. B \textbf{86}, 184516 (2012).
\bibitem{Zhang} F. Zhang, C. L. Kane, and E. J. Mele, Phys. Rev. Lett. \textbf{111}, 056402 (2013).
\bibitem{Nagaosa2} S. Nakosai, J. C. Budich, Y. Tanaka, B. Trauzettel, and
N. Nagaosa, Phys. Rev. Lett. \textbf{110}, 117002 (2013).
\bibitem{Gong} Z. Gao and W. J. Gong, Phys. Rev. B \textbf{94}, 104506 (2016).
\bibitem{Liuxj} X. J. Liu, C. L. M. Wong, and K. T. Law, Phys. Rev. X \textbf{4}, 021018 (2014).
\bibitem{Qixl} S. B. Chung, J. Horowitz, and X. L. Qi, Phys. Rev. B \textbf{88}, 214514 (2013).

\bibitem{BDI} S. V. Aksenov, A. O. Zlotnikov, and M. S. Shustin, Phys. Rev. B \textbf{101}, 125431 (2020).
\bibitem{BDI1} A. P. Schnyder, S. Ryu, A. Furusaki, and A. W. W. Ludwig, Phys. Rev. B \textbf{78}, 195125 (2008).
\bibitem{BDI2} X.-L. Qi, T. L. Hughes, and S.-C. Zhang, Phys. Rev. B \textbf{82}, 184516 (2010).
\bibitem{BDI3} J. J. He, J. Wu, T.-P. Choy, X.-J. Liu, Y. Tanaka, and K. T. Law, Nat. Commun. \textbf{5}, 3232 (2014).
\bibitem{BDI4} A. Ueda and T. Yokoyama, Phys. Rev. B \textbf{90}, 081405(R) (2014).
\bibitem{dimerized kitaev1} R. Wakatsuki, M. Ezawa, Y. Tanaka, and N. Nagaosa, Phys. Rev. B \textbf{90}, 014505 (2014).
\bibitem{dimerized kitaev2} M. A. R. Griffith, E. Mamani, L. Nunes, and H. Caldas, Phys. Rev. B \textbf{101}, 184514 (2020).
\bibitem{dimerized kitaev3} M. Ezawa, Phys. Rev. B \textbf{96}, 121105 (2017).
\bibitem{long range1} D. Vodola, L. Lepori,  E. Ercolessi, A. V. Gorshkov, and G. Pupillo, Phys. Rev. B \textbf{113}, 156402 (2014).
\bibitem{long range2} A. Alecce and L. Dell'Ann, Phys. Rev. B \textbf{95}, 195160 (2017).
\bibitem{long range3} A. Dutta and A. Dutta, Phys. Rev. B \textbf{96}, 125113 (2017).
\bibitem{recursive1} C. H. Lewenkopf, and E. R. Mucciolo, J. Comput. Elecron. \textbf{12}, 203 (2013).
\bibitem{recursive2} Y. Peng, Y. M. Bao, and F. von Oppen, Phys. Rev. B \textbf{95}, 235143 (2017).
\bibitem{recursive3} A. Li, A. Yamakage, K. Yada, M. Sato, and Y. Tanaka, Phys. Rev. B \textbf{86}, 174512 (2012).
\bibitem{disorder1} C. B. Hua, R. Chen, D. H. Xu, and B. Zhou, Phys. Rev. B \textbf{100} 205302 (2019).
\bibitem{disorder2} V. M. Martinez Alvarez and M. D. Coutinho-Filho, Phys. Rev. A \textbf{99}, 013833 (2019).
\bibitem{disorder3} A. R. Akhmerov, J. P. Dahlhaus, F. Hassler, M. Wimmer, and C. W. J. Beenakker, Phys. Rev. Lett. \textbf{106}, 057001 (2011).


\end{thebibliography}
\end{document}